\documentclass[twocolumn,showpacs,amsmath,amssymb]{revtex4}

\usepackage{graphicx}
\usepackage{times}
\usepackage{bm}

\begin{document}
\preprint{L05-1285, 001529APL}

\title{Electron transport in ZnO thin films}

\author{T. Makino$^{\textbf{a)}}$} \altaffiliation[\textbf{a):} Present address: ]{Graduate School of Material Science, University of Hyogo, Kamigori-cho, Ako-gun, Hyogo 678-1297, Japan} \email[electronic mail: ]{tmakino@riken.jp}

\author{Y. Segawa}
\affiliation{Photodynamics Research Center, Institute of Physical and Chemical 
Research (RIKEN), Aramaki aza Aoba 519-1399, Aobaku, Sendai 980-0845, 
Japan}

\author{A. Tsukazaki} \author{A. Ohtomo} \author{M. Kawasaki$^{\textbf{b)}}$} \altaffiliation[\textbf{b)}: Also at: ]{Combinatorial Exploration Material Science and Technology, Tsukuba, Japan}

\affiliation{Institute for Materials Research, Tohoku University, 2-1-1 Katahira, Aoba, Sendai, 980-8577, Japan}

\date{\today}
\begin{abstract}
Epitaxial, \textit{n}-type ZnO films grown by a laser molecular-beam epitaxy method
were investigated by the temperature-dependent Hall-effect technique.
The 300-K carrier concentration and mobility were about $n_s \sim 10^{16}$~cm$^{-3}$ and 440~cm$^{2}$/Vs, respectively.
Transport characteristics are
calculated by solving the Boltzmann transport equation using a variational method. Mobility limit of 430~cm$^{2}$/Vs
was calculated at 300~K. The temperature dependence of the mobility for an undoped film is calculated and agrees favorably well with experimental data if physical parameters are
chosen so as to approach those.
In the experimental `mobility versus concentration' curve, unusual phenomenon was observed, i.e., mobilities at $n_s \sim 5\times$ 10$^{18}$ cm$^{-3}$ are significantly smaller than those at higher densities above $\sim 10^{20}$ cm$^{-3}$. 
It is qualitatively explained in terms of electron-plasmon interaction.
\end{abstract}
\maketitle

\newpage
ZnO is direct band gap semiconductor suitable for the development 
of efficient blue and UV optoelectronic devices. Attempts have extensively 
been made to incorporate shallow acceptor impurities in this 
material. We have recently developed a reliable and reproducible 
method to grow p-type ZnO layers, leading to the observation of electro-luminescence 
from a homoepitaxial p-n junction structure~\cite{nature_mat_tsukazaki}. The exploit for discovering 
the p-type conduction has noticed us the importance of quality 
improvement of undoped epilayers as a starting point of reliable 
doping~\cite{tsukazaki1}. In this work, the electrical properties of epilayers 
are investigated since mobility is considered to be the \textit{figure 
of merit} for material characterization.

Several investigators have reported the mobilities in epitaxial ZnO layers grown by various techniques~\cite{miyamoto_mob1,ellmer_mob1,ginley_mrs,kato_jjap_42_2241,iwata_pssa_180_287,kaidashev:3901}. The room-temperature values are relatively poor compared with bulk ZnO. Our recent mobilities~\cite{nature_mat_tsukazaki} surpassed, however, the best bulk data ever reported~\cite{dclook_mob1,dclook_mob2,rode_bkmob1} and are significantly 
higher than the calculation results performed for bulk crystals 
by Rode (p.~49 of Ref.~\onlinecite{rode_bkmob1}). Since the calculated result presents the material limit~\cite{rode_bkmob1,albrecht:6864}, it needs
to calculate the mobility again with a different set of parameters so as to increase its limit.
The corresponding critical discussion of their choice will be made in this work.

To date, there have only been a few attempts~\cite{ellmer_mob1,miyamoto_mob1} to model electron 
transport in \textit{doped} ZnO. The present publication also reports carrier concentration dependence 
of mobility including an effect of the electron-plasmon interaction~\cite{fischetti_mob1,lowney_mob1}.

Table~I gives the ZnO materials parameters that are used in these 
transport theory calculation~\cite{lowney_mob1,seeger_bkmob1,LBZincoxide}.
Since the current is perpendicular to the \textit{c}-axis in our case~\cite{rode_bkmob1}, the piezoelectric coefficient of 0.21 was used, the definition of
which is found in p.~36 of Ref.~\onlinecite{rode_bkmob1}.
We have not adopted the relaxation
time approximation for the mechanisms that involve relatively 
high-energy transfers, \textit{e.g.}, longitudinal optical 
phonons. The polar optical-phonon energy, $\hbar \omega_{LO}$, is 72~meV, and corresponding
Debye temperature (837~K) by far exceeds the room temperature~\cite{LBZincoxide}. The pair of $\epsilon_0$ and $\epsilon_\infty $ was selected from the literature as to make their difference smaller. The effective band mass was adopted instead of its polaron mass. Otherwise, the resulting mobility become smaller. Since Rode's iterative technique takes a long time to reach its convergence~\cite{rode_bkmob1,fischetti_mob1}, the present computations 
are based on the variational principle method~\cite{lowney_mob1,ruda_mob1}. This method allows
the combination of all scattering mechanism without invoking
Mattheissen's rule.

The following electron scattering mechanisms are considered: (1)
polar optical-phonon scattering, (2) ionized-impurity 
scattering, (3) acoustic-phonon scattering through the deformation 
potentials, and (4) piezo-electric interactions~\cite{seeger_bkmob1,lowney_mob1,ruda_mob1}. The final expression 
used is as follows~\cite{lowney_mob1}:

\begin{equation}
\mu =22.5 \left[F_{1/2}(\eta ) T^{1/2}  \left(\frac{{m^*}}{{m_0}}\right)^{3/2} z_l \left(\frac{1}{{{\epsilon }_{\infty }}}-\frac{1}{{{\epsilon}_0}}\right)\right]^{-1} \frac{\bm{D}_{3/23/2}}{\bm{D}}.
\end{equation}

The terms $\epsilon_0$ and $\epsilon_\infty $
refer to the low- and high-frequency dielectric constants, respectively. \textit{m}$^{{*}}$ 
refers to the electron effective mass in units of the free electron 
mass in rest. \textit{T} refers to the temperature, and ${z}_{l}$ is 
the reduced polar optical phonon energy defined as $\hbar \omega_{LO}/k T$. $F_{1/2}({\eta})$ is 
the half-integral Fermi-Dirac integral and $\eta$ is 
the reduced Fermi energy given in thermal units. The terms $\bm{D}_{3/23/2}$ 
and $\bm{D}$ are defined and explained in Ref.~\onlinecite{lowney_mob1}.

All films were grown on insulating and lattice-matched ScAlMgO$_{4}$ (SCAM) 
substrates by laser molecular-beam epitaxy. ZnO single-crystal 
and (Ga,Zn)O ceramics targets were ablated by excimer laser pulses 
with an oxygen flow of $1\times 10^{-6}$~torr. We have grown them at about 
1000 $^\circ$C with introduction of semi-insulating (Mg, Zn)O buffer 
layers. The detailed description of fabrication methods has been given elsewhere~\cite{ohtomo_sst,makino_sst}. The films were patterned into Hall-bars and the contact 
metal electrodes were made with Au/Ti for \textit{n}-type film, giving good 
ohmic contact. Carrier concentrations $n_s$ and Hall mobilities 
$\mu $ were measured at various temperatures.

Figure~1 shows temperature dependence of mobility ($\mu $) for an undoped ZnO film and for a bulk crystal by closed and open circles. The values of $\mu $ are 440~cm$^{2}$/Vs and 5,000~cm$^{2}$/Vs at 300 and 100~K, respectively. Below 
100~K, the distribution of activated carriers in this sample 
is too sparse to detect the sizable Hall electromotive force.
We did not observe 
abnormal temperature dependence requiring a so-called two-layer 
analysis in these data~\cite{look_apl_70_3377}. It thus suggests the negligible contribution
from a degenerate interfacial layer to our samples probably because ours were grown on lattice-matched substrates.

\begin{figure}[htbp]
	\includegraphics[width=0.9\linewidth]{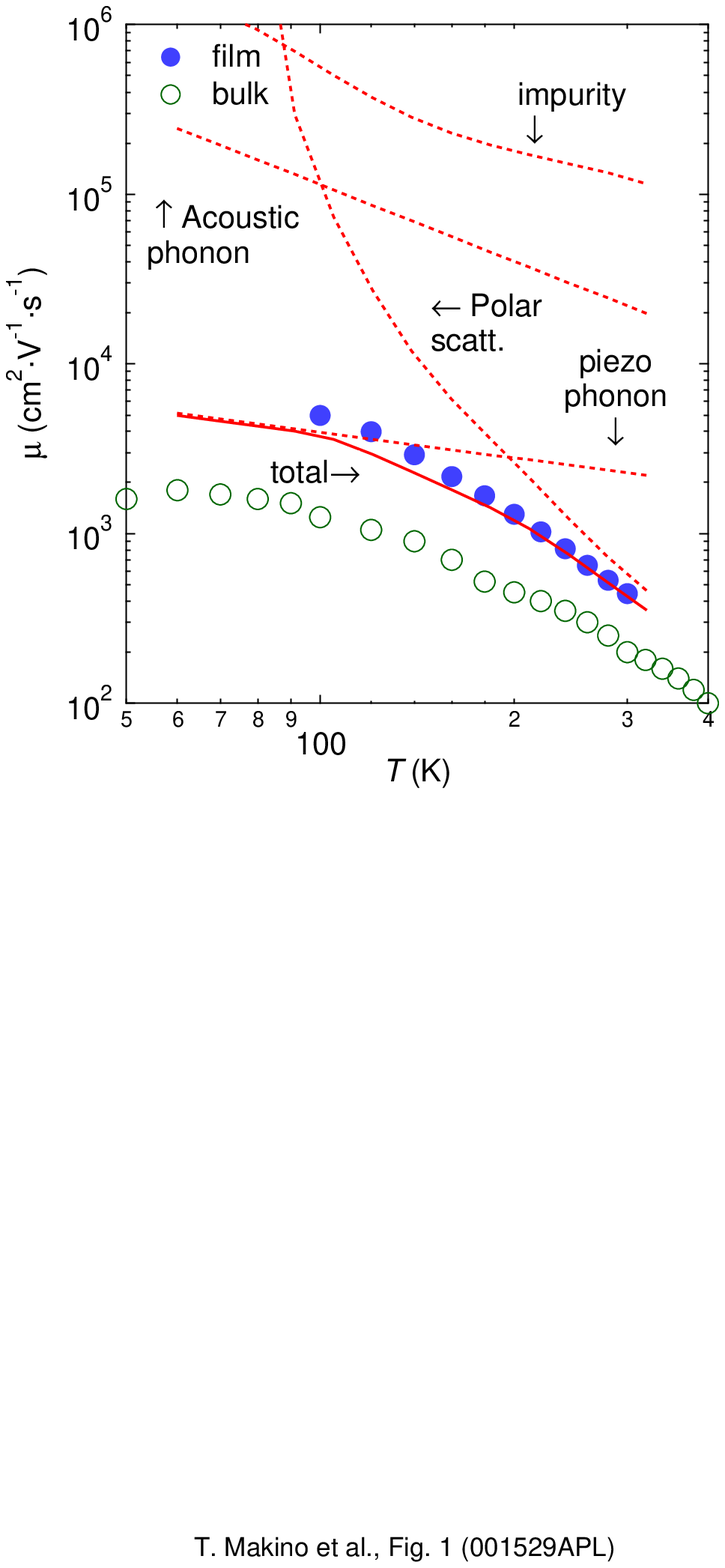}
	\caption{(Color online). Components (dashed curves) of the total electron mobility (solid line) at nondegenerate 
case versus temperature in ZnO. Calculations are for nondegenerate 
conditions. Also shown by closed and open circles are experimental data from
an undoped film and a bulk crystal (cited from Ref.~10).}
	\label{fig1}
\end{figure}
We have derived partial mobilities by accounting for respective 
scattering mechanisms in the nondegenerate (undoped) limit. The 
results are given in Fig.~1 as a function of temperature by dashed
lines. The curve for ionized impurity scattering are calculated
assuming that the material is uncompensated and concentration
being equal to $n_s$. The carrier concentration data have been given
elsewhere~\cite{nature_mat_tsukazaki}; data were fitted with
activation energies of $\sim $30 and 60~meV. The total electron mobility 
calculated by combining all of the partial scattering mechanisms 
is also given by a solid line. The
relative importance of the various scattering mechanisms mimics
that found in ZnS~\cite{ruda_mob1}. For example, the polar optical phonon scattering
at higher temperatures controls inherent mobility limit curve.
At low temperatures (down to 100~K) on the other hand, piezoelectric
phonon interaction is the dominant scattering mechanism. Our experimental data are in reasonably good
agreement with theory. The mobility limit at 300~K is about 430~cm$^{2}$/Vs. The probable discrepancy between them is because our 
model calculates drift mobility, while experimental results correspond 
to Hall mobility. The theoretical Hall factor was given in Ref.~\onlinecite{rode_bkmob1}.

\begin{figure}[htbp]
	\includegraphics[width=0.9\linewidth]{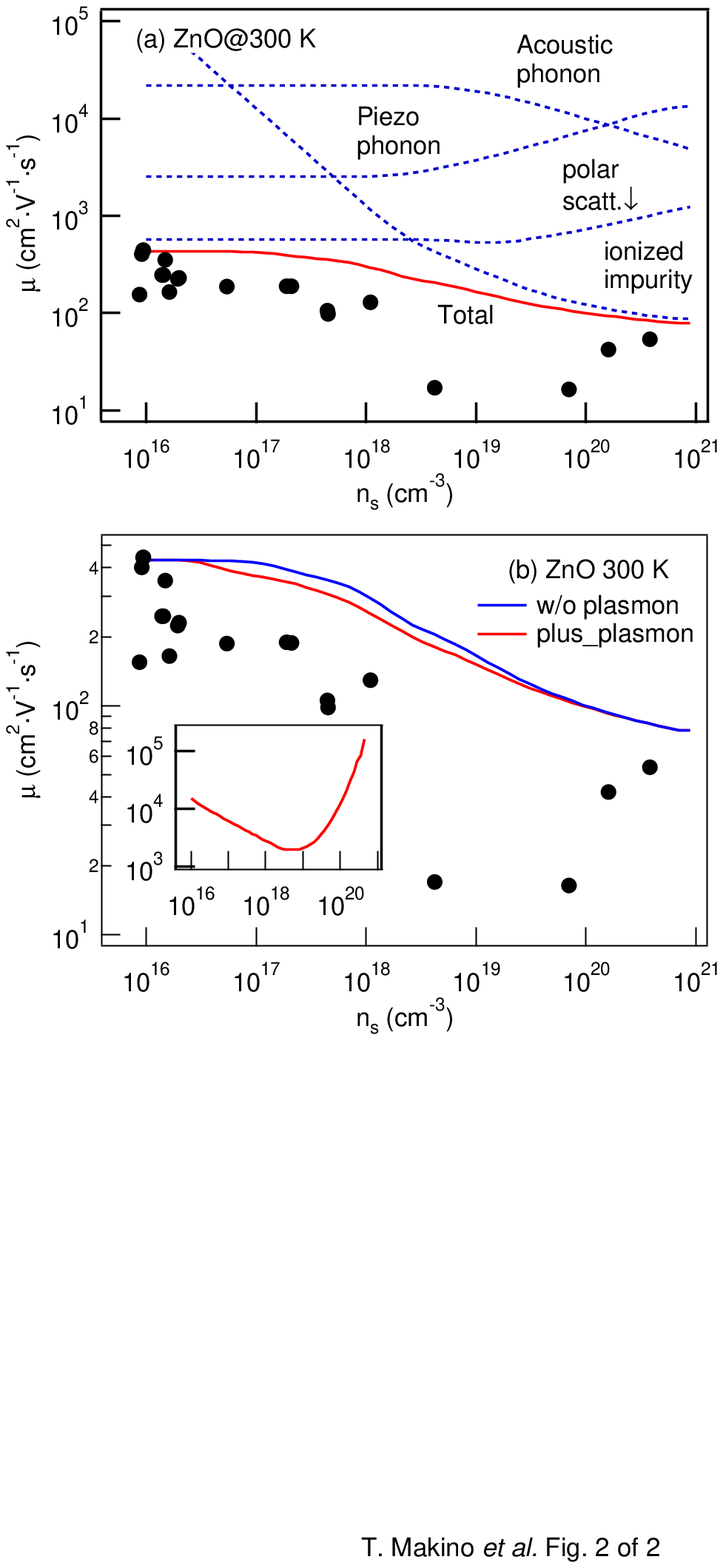}
	\caption{(a) Comparison of drift mobility calculations (solid 
curve) with the Hall effect measurements for undoped and doped 
epitaxial films (filled circles). The contributions of various scattering 
mechanisms to the total mobility are also shown by dashed curves. (b) Total theoretical mobility with (lower, full line)
and without (upper, dashed line) including the effect of electron-plasmon interaction. Also shown in the inset is the partial mobility.}
	\label{fig2}
\end{figure}
On the other hand, the situation is somewhat different for the 
cases of Ga-doped \textit{n}-type films~\cite{makino_int_Ga}. Figure~2(a) shows 300-K experimental 
mobilities plotted against carrier concentration. The mobilities 
of doped films are significantly smaller than those of undoped 
one~\cite{makino18,makino19}. Obviously, this tendency can be qualitatively attributed to the increased density of impurities. For quantitative comparison, 
partial mobilities are calculated and given in Fig.~2(a) by dashed
lines. We have taken the effects of screening for both ionized 
impurities and polar optical phonon scattering into account. Polar interactions reflecting the ionicity of
the lattice are dominant in scattering mechanism, while, at heavier doping 
levels, ionized impurity scattering controls the inherent mobility 
limit curve~\cite{ruda_mob1,lowney_mob1}. The experimental data agree well with our calculation (solid curve)
except for the intermediate concentration range. Particularly, our model could not reproduce 
a relative minimum in the ``$\mu $ versus $n_s$'' curve, which 
has been experimentally observed in this intermediate doping 
range. The situation 
at $n_s$ \texttt{>} 10$^{20}$ cm$^{-3}$ could be improved probably if the 
effects of non-parabolic band structure as well as of clustering 
of charged carriers would be taken into account~\cite{ellmer_mob1}, which is beyond 
the scope of our work.

Fischetti~\cite{fischetti_mob1} and Lowney and Bennett~\cite{lowney_mob1} obtained theoretical $\mu $ curve having a ``dip'' in the intermediate concentrations 
by including a higher-order effect, i.e., plasmon scattering. 
Plasmons are the collective excitations of free electrons against 
the background charges. It has been shown that if decay due to 
collisional damping dominates over Landau damping, the electron-plasmon 
scattering can reduce somewhat the mobility for electron concentrations 
greater than 10$^{17}$~cm$^{-3}$. The coupling of the plasmon to the 
polar optical-phonon modes have been neglected because there 
is no sufficiently tractable theory for the hybrid modes, and 
because such coupling should not have much of an effect on the overall numerical 
results for the mobility. We thus decide to deal with the pure uncoupled 
modes throughout.

The momentum relaxation time $\tau_{pl}$ for an electron to absorb or emit a plasmon of energy $\hbar \omega_{pl}=n_s e^2/\epsilon_0 m^*$ is given by; 
\begin{equation}
\frac{1}{{{\tau }_{pl}}}=\Theta ({E^\prime}) \frac{{{\omega }_{pl}}{k^\prime}}{{a_0}}\frac{1-{f_0}({E^\prime})}{1-{f_0}(E)}  \left({N_p}+\frac{1}{2}\pm \frac{1}{2}\right)\times \int _{{X_c}}^{1}\left(1-\frac{{k^\prime}}{k}X\right)\frac{1}{{q^2}}dX
\end{equation}
where upper and lower signs represent absorption and emission, 
respectively. $X_c=[1+\Psi-(q_c/k)^2/2 \Psi^{1/2}]$, 
$q_c$ is the plasmon cut-off wavenumber, above which plasma oscillation 
cannot be sustained. The cut-off wavenumber is on the order of 
the inverse Thomas-Fermi screening length defined elsewhere~\cite{lowney_mob1}. 
$E^\prime = E \pm \hbar \omega_{pl}$, and $N_p$ is the Bose occupation number of plasmons and $\Psi=(1 \pm \hbar \omega_{pl}/E)$. The relaxation time 
approximation was adopted here because it is valid for the entire 
concentration range of interest according to Ref.~\onlinecite{fischetti_mob1}.

The dependence of the cut-off wavenumber and plasma frequency 
on doping density leads to interesting phenomena for plasmon 
scattering such as a relative minimum in mobility as a function 
of concentration. Figure~2(b) shows the effect of electron-plasmon 
interaction on the mobility. The uppermost curve is drawn without including 
this effect, while the next one takes it into account. Notice, 
in particular, the ``dip'' is at $n_s \simeq 5\times $
10$^{18}$ cm$^{-3}$. The partial mobility is shown in the inset.
Our theory thus describes a rise due to the increase
in the plasmon frequency at these doping levels as has been observed 
experimentally, although there is poor quantitative agreement 
between theory and experiment. We propose that more experiments 
using further optimized samples be done in this doping range 
to determine if the tendency on mobility agrees not only qualitatively but also quantitatively.

In summary, electron transport characteristics of ZnO are experimentally 
studied and calculated by solving the Boltzmann transport equation 
exactly for all major scattering mechanisms. The polar optical 
phonon and ionized impurity electron scattering mechanisms are 
properly screened in this treatment. Temperature dependence of 
the mobility is investigated for nondegenerate samples, and compare 
with our experimental data. Reasonable agreement between experimental 
data and theory was found. Mobility limits are derived for 
this material as 430~cm$^{2}$/Vs at 300~K. Electron mobility was also calculated 
as a function of carrier concentration. Even if we take the effect 
of electron-plasmon interaction into consideration, our experimental 
mobilities for doped films are significantly lower than theoretical 
values especially for the concentration range around 10$^{19}$ cm$^{-3}$.

\textit{Acknowledgement} --- One of the authors (T. M.) thanks H. S. Bennett of NIST, D. L. 
Rode of Washington University, St. Louis, H. Ruda of University 
of Toronto, Canada, and B. Sanborn of Arizona State University for helpful 
discussions. Thanks are also due to Shin Yoshida for technical
assistance during our experiments. 

\newpage

\newpage

\begin{table}[htbp]
	\caption{List of material parameters. (Ref.~\onlinecite{LBZincoxide})}
	\label{tbl1}
	\begin{ruledtabular}
		\begin{tabular}{ccc}
Parameter & Symbol & Value\\
\hline

Energy gap (300 K)  & \textit{E}$_{g}$ (eV) & 3.37\\
Electron effective mass & m$^*$ ($m_0$) & 0.24\\
High-freq. dielectric constant & $\epsilon_\infty$ & 8.1\\
Static dielectric constant & $\epsilon_0$ & 4.0\\
Optical phonon energy & $\hbar \omega_{LO}$ (meV) & 72\\
Acoustic deformation potential & $E_d$ (eV) & 3.8\\
Longitudinal elastic constant & $c_l$ (10$^{7}$ dyn/cm$^{2}$) & 2.05\\
Piezoelectric coefficient  & $P_\perp $ & 0.21\\
		\end{tabular}
	\end{ruledtabular}
\end{table}
\end{document}